\documentclass[aps,twocolumn,floatfix, showpacs, superscriptaddress]{revtex4-1}
\usepackage[pdftex]{graphicx}
 \usepackage{amsmath}
\usepackage[T1]{fontenc}
\usepackage{amssymb}
\usepackage{amsfonts}
\usepackage{flushend}
 \usepackage{amsmath} 
\usepackage{lipsum}
\usepackage{amsfonts} 
\usepackage{amssymb, mathrsfs}
\usepackage{braket}
\usepackage{graphicx} 
\usepackage{subfigure}
\usepackage{bbm}
\def\beq{\begin{equation}}
\def\eeq{\end{equation}}
\def\bsp{\begin{split}}
\def\esp{\end{split}}
\def\bea{\begin{eqnarray}}
\def\eea{\end{eqnarray}}
\def\ba{\begin{array}}
\def\ea{\end{array}}

\def\dg{\dagger}

\def\lb{\left(}
\def\rb{\right)}

\def\l.{\left.}
\def\r.{\right.}

\def\ra{\rangle}
\def\la{\langle}

\def\bo{\bold{k}}

\begin{document}

\date{\today}
\title{Topological Magnon Insulator in NonCoplanar Kagom\'e Antiferromagnets: Supplemental Material of  arXiv:1608.04561}
\author{S. A. Owerre}
\affiliation{Perimeter Institute for Theoretical Physics, 31 Caroline St. N., Waterloo, Ontario N2L 2Y5, Canada.}
\affiliation{African Institute for Mathematical Sciences, 6 Melrose Road, Muizenberg, Cape Town 7945, South Africa.}

\begin{abstract}
Topological magnon insulators in insulating Kagom\'e ferromagnets have been extensively studied in a series of papers. It has been established that Dzyaloshinskii-Moriya interaction (DMI) is the key ingredient to observe a nontrivial topological magnon with edge modes. However, 
insulating  antiferromagnets on the Kagom\'e lattice are frustrated systems considered as a playground for studying quantum spin liquid physics. In these systems the DMI  can induce a coplanar but noncollinear magnetic orders with a $\mathbf{q}=0$ propagating wavevector.  We show that  topological magnon bands are absent in this coplanar spin texture  in sharp contrast to collinear ferromagnets with DMI. Hence,  geometrically frustrated Kagom\'e  antiferromagnets can be deemed topologically trivial. The presence of an out-of-plane magnetic field in these frustrated magnets  induces noncoplanar spin textures exhibiting  a nonzero spin scalar chirality.  We show that the field-induced spin  chirality provides topological magnon bands in Kagom\'e antiferromagnets without the need of DMI and survives in the chiral spin liquid phase of frustrated magnets. Possible experimentally material includes  iron jarosite KFe$_3$(OH)$_{6}$(SO$_{4}$)$_2$.  \end{abstract}
\pacs{73.43.Nq, 66.70.-f, 75.10.Jm}
\maketitle

\section{Introduction}
 In recent years,  the concept of topological band theory has been extended to nonelectronic systems such as  magnons \cite{alex1, alex0, alex2,alex5,alex4, sol1,sol,sol2, alex5a, alex6,  sol4,  mok,su,fyl,lifa} and phonons \cite{pho,pho1,pho2,pho3,pho4,pho5}.   In the former, the DMI \cite{dm} is the primary source of  topological magnon bands  and magnon edge states \cite{lifa,alex4}, as well as thermal magnon Hall effect \cite{alex0,alex1}.  These systems are dubbed  topological magnon insulators \cite{lifa} in analogy to topological insulators in fermionic systems with spin-orbit coupling (SOC) \cite{guo,kane,kane1}. However, as magnons are charge-neutral quasiparticles, the perfectly conducting  edge states are believed to be useful for dissipationless transport applicable to magnon spintronics. Our conception of DMI being the primary source of topological effects in magnonic systems has been firmly established  by recent experimental realization of topological magnon insulator in collinear Kagom\'e ferromagnet Cu(1-3, bdc) \cite{alex5a}. In these systems the DMI  comes naturally because the Kagom\'e lattice lacks an inversion center.
 
  In reality, however, there are more frustrated Kagom\'e magnets than collinear ferromagnets. The physics associated with the  former has no analogy with the latter. The former are considered as a candidate for quantum spin liquid physics due to an extensive classical degeneracy that prevent magnetic ordering to lowest accessible temperatures.  However, in recent experimental synthesis it has been shown that the effects of SOC or DMI are not negligible in most Kagom\'e antiferromagnets. The DMI appears as a perturbation to the Heisenberg spin exchange. One of the striking  features of this perturbation in frustrated  Kagom\'e systems is that it induces magnetic ordering with a $\bold q=0$ propagation wavevector.  Thus, it suppresses the spin liquid phase of Kagom\'e antiferromagnets up to a critical value \cite{sup1, men3}. Various experimentally accessible  frustrated  Kagom\'e antiferromagnets show evidence of coplanar/noncollinear $\bold q=0$ magnetic ordering at specific temperatures. The famous one is iron jarosite KFe$_3$(OH)$_{6}$(SO$_{4}$)$_2$ \cite{sup1a,sup2}. The unanswered question is how does topological effects arise in these systems? From the experimental perspective, it has been previously  shown that iron jarosite posseses a finite spin scalar chirality induced by an out-of-plane magnetic field but no topological properties were measured \cite{sup1a}.  In a recent Letter \cite{me}, we have provided evidence of  magnon Hall effect and   thermal Hall conductivity $\kappa_{xy}$ for this Kagom\'e material, which originates from non-coplanar spin texture and survives in the absence of DMI and magnetic ordering. Topological magnon insulator with magnon edge modes are another area of recent development \cite{alex5a, lifa, alex4}.
     
 In this report, we complete our analysis of topological magnon effects in geometrically frustrated Kagom\'e antiferromagnets by providing evidence of topological magnon edge modes. The main purpose of this report is   to relate finite thermal Hall conductivity to topologically protected magnon edge states. We consider three different models, viz:   $(i)$ bilayer Kagom\'e ferromagnets coupled antiferromagnetically or layer antiferromagnets. $(ii)$ The model Hamiltonian for iron jarosite KFe$_3$(OH)$_{6}$(SO$_{4}$)$_2$ with DMI and second nearest neighbour interaction \cite{sup2}.  $(iii)$ The XXZ model for Kagom\'e antiferromagnet without the DMI.  The arrangement of this paper is as follows. In Sec. II we introduce the three models, analyze the magnon tight binding models, and present the distinctive magnon bands. Sec. III introduces the concept of topologically protected magnon edge modes and fictitious Chern numbers. We relate these concepts to experimentally accessible thermal magnon Hall effect. In Sec. IV we conclude and discuss potential methods to realize topological magnon bands in geometrically frustrated Kagom\'e antiferromagnets with/without DMI.

\section{  Model Hamiltonian }
\subsection{Model I}
In various frustrated Kagom\'e magnets, a strong out-of-plane magnetic field is sufficient to circumvent frustrated interactions and leads to magnetic ordering. A good example is the bilayer frustrated Kagom\'e magnet Ca$_{10}$Cr$_7$O$_{28}$ \cite{Balz},  which shows evidence of ferromagnetic alignment at a magnetic field of magnitude $h\sim 11~\text{Tesla}$. The frustrated  Kagom\'e volborthite Cu$_3$V$_2$O$_7$(OH)$_2$ $\cdot$2H$_2$O also shows evidence of magnetic ordering  at several field values \cite{Yo,Yo1}. In the ordered regimes, the magnetic excitations are definitely magnons. Owing to the fact that many Kagom\'e magnets come naturally in bilayer forms,    we first consider bilayer  Kagom\'e   magnets with non-negligible interlayer coupling. We assume that the top layer is placed right above the bottom layer forming  AAA-stacked pattern. The Hamiltonian is given by \begin{align}
\mathcal H&= \sum_{\la i, j\ra\tau} \lb \mathcal{J}{\bf S}_{i}^\tau\cdot{\bf S}_{j}^\tau + \boldsymbol{\mathcal D}_{ij}\cdot{\bf S}_{i}^\tau\times{\bf S}_{j}^\tau\rb -h\hat{\bold z}\cdot\sum_{i\tau} {\bf S}_{i}^\tau\label{h}\\&\nonumber + \mathcal J_t\sum_{i}{\bf S}_{i}^t\cdot{\bf S}_{i}^b,
\end{align}
where ${\bf S}_{i}$ is the spin moment at site $i$, $\tau$  labels the top ($t$) and bottom ($b$) layers respectively,  $h$ is an external magnetic field in units of $g\mu_B$. We consider the case of ferromagnetic intra-layer exchange $\mathcal J<0$ and antiferromagnetic interlayer coupling $\mathcal J_t>0$ with an out-of-plane magnetic field $h$.   At zero magnetic field, the spins on the top and bottom layers lie in opposite directions on the $x$-$y$ Kagom\'e planes, and interlayer coupling is antiferromagnetic.  A nonzero magnetic field is expected to introduce canting along the direction of the field. Hence, the ground state is no longer the collinear ferromagnets.

 In the classical limit, the spin operators   can be represented as classical vectors,  written as
 $\bold{S}_\tau= S\bold{n}_\tau$, where $\bold{n}_\tau=\lb\sin\chi\cos\theta_\tau, \sin\chi\sin\theta_\tau,\cos\chi \rb$
 is a unit vector. The classical energy gives

\begin{align}
e_{cl}=-|\mathcal J|+\frac{\mathcal J_t}{2}\cos2\chi- h\cos\chi,
\end{align}
 where  $e_{cl}=E_{cl}/6NS^2$, $N$ is the  number of sites per unit cell, and the magnetic field is rescaled in units of $S$. Minimizing the classical energy yields the canting angle $\cos\chi= h/(h_s= 2 \mathcal J_t )$. We see that both the out-of-plane DM vector ($\boldsymbol{\mathcal D}_{ij}=\mathcal D\hat{\bold z}$) and the in-plane DM vector ($\boldsymbol{\mathcal D}_{ij}=\mathcal D\hat{\bold x}$)  does not contribute to the classical energy due to ferromagnetic ordering on each layer.

For the magnon excitations above the classical ground state, the basic procedure involves rotating the spins from laboratory frame to local frame  by the spin oriented angles $\theta_\tau$  about the $z$-axis. Due to the field-induced canting, we perform another rotation about the $y$-axis with the canting angle $\chi$. The total rotation matrix is given by 
\begin{align}
\mathcal{R}_z(\theta_\tau)\cdot\mathcal{R}_y(\chi)
=\begin{pmatrix}
\cos\theta_\tau\cos\chi & -\sin\theta_\tau & \cos\theta_\tau\sin\chi\\
\sin\theta_\tau\cos\chi & \cos\theta_\tau &\sin\theta_\tau\sin\chi\\
-\sin\chi & 0 &\cos\chi
\end{pmatrix},
\label{rot}
\end{align}
where $\theta_\tau$ labels the spin oriented angles on each layer with $\theta_t=\pi$ for the top layer, $\theta_b=0$ for the bottom layer, and $\chi$ is the  field canting angle. Hence, \bea \bold{S}_i=\mathcal{R}_z(\theta_\tau)\cdot\mathcal{R}_y(\chi)\cdot\bold S_i^\prime,\eea which can be written explicitly as 
\begin{align}
&S_{i\tau}^x=\pm S_{i\tau}^{\prime x}\cos\chi \pm S_{i\tau}^{\prime z}\sin\chi,\label{trans}\nonumber\\&
S_{i\tau}^y=\pm S_{i\tau}^{\prime y},\\&\nonumber
S_{i\tau}^z=- S_{i\tau}^{\prime x}\sin\chi + S_{i\tau}^{\prime z}\cos\chi,
\end{align}
where $-(+)$  applies to the layers $t(b)$ respectively. This rotation does not affect the ferromagnetic $\mathcal J$-term on each layer. A crucial distinguishing feature of this system is that  both the out-of-plane DMI ($\boldsymbol{\mathcal D}_{ij}=\mathcal D\hat{\bold z}$) and the in-plane DMI ($\boldsymbol{\mathcal D}_{ij}=\mathcal D\hat{\bold x}$) contribute in the present system. This is as a result of field-induced canting of the system. It should be noted that this is not the case in previously studied  Kagom\'e ferromagnets  \cite{alex1, alex0, alex2,alex5,alex4, sol1,sol,sol2, alex5a, alex6,   lifa}.  To linear order in $1/S$ expansion  we have
\begin{align}
&\mathcal H_{DM,z}= \mathcal D\cos\chi\sum_{\la i, j\ra\tau}\hat{\bold z}\cdot {\bf S}^{\prime\tau}_{i}\times {\bf S}_{j}^{\prime\tau} +\mathcal{O}(1/S),\\&
\mathcal H_{DM,x}=  \sigma \mathcal D\sin\chi\sum_{\la i, j\ra\tau}\hat{\bold z}\cdot {\bf S}_{i}^{\prime\tau}\times {\bf S}_{j}^{\prime\tau} +\mathcal{O}(1/S),
\end{align}
  where $\sigma=\mp$ for top and bottom layers respectively. We see that the latter case has opposite signs on both layers.  
  
 Next, we study the excitations above the classical ground state by using the Holstein-Primakoff spin bosonic operators for the rotated prime coordinates:   $S_{i,\tau}^{\prime x}=\sqrt{S/2}(b_{i,\tau}^\dg+b_{i,\tau})$, $S_{i,\tau}^{\prime y}=i\sqrt{S/2}(b_{i,\tau}^\dg-b_{i,\tau})$, and $S_{i,\tau}^{\prime z}=S-b_{i,\tau}^\dg b_{i,\tau}$. The  Hamiltonian maps to a magnon tight binding model\begin{align}
\mathcal H_{SW}&=v_0\sum_{i\tau} n_{i\tau} -v_D\sum_{\la ij\ra\tau} \lb e^{-i \phi_{ij}}b_{i\tau}^\dagger b_{j\tau} +h.c.\rb \label{hpp3}\\&\nonumber-v_t\sum_{i\in \tau, j\in \tau^\prime}\bigg[(n_{i\tau}+ n_{j\tau^\prime})\cos2\chi \\&\nonumber+( b_{i\tau}^\dagger b_{j\tau^\prime}+ h.c.)\cos^2\chi-( b_{i\tau}^\dagger b_{j\tau^\prime}^\dagger+ h.c.)\sin^2\chi\bigg],
\end{align}
where $n_{i\tau}=b_{i\tau}^\dagger b_{i\tau}$ is the occupation number, $v_0= 4v_{J} +h\cos\chi,~v_t=\mathcal J_t S,~v_J=|\mathcal J|S$, and $v_D= S\sqrt{\mathcal J^2 +\mathcal  D_{x,z}^2}$ with  $\mathcal D_{z}=\mathcal D\cos\chi$, $\mathcal D_{x}=\sigma\mathcal D\sin\chi$. The fictitious magnetic flux on each triangle of the Kagom\'e lattice is given by $\phi=\arctan (\mathcal D_{x,z}/|\mathcal J|)$.  Using the vectors $\Psi_\bo^\dg=(\psi_\bo^\dagger, ~\psi_{-\bo})$, with $\psi^\dagger_{\bold k}= (b_{\bold{k}A}^{\dagger},\thinspace b_{\bold{k} B}^{\dagger},b_{\bold{k} C}^{\dagger},\thinspace b_{\bold{k} A^\prime}^{\dagger},\thinspace b_{\bold{k} B^\prime}^{\dagger},\thinspace b_{\bold{k} C^\prime}^{\dagger})$, the momentum space Hamiltonian is given by $\mathcal H_{SW}=\frac{1}{2}\sum_{\bold k}\Psi^\dagger_{\bold k}\cdot \boldsymbol{\mathcal{H}}_{AFM}(\bold k)\cdot\Psi_{\bold k},$ where
 \begin{align}
\boldsymbol{\mathcal{H}}_{AFM}(\bold k)=\left(
\begin{array}{cc}
\bold  A(\bo,\phi)& \boldsymbol{{B}}\\
 \boldsymbol{{B}}&  \bold A^*(-\bo,\phi)
\end{array}
\right),
\end{align}

\begin{align}
\boldsymbol A(\bo,\phi)&= 
\begin{pmatrix}
\bold a(\bo,\phi)& \boldsymbol{{b}}\\
\boldsymbol{{b}} &\bold a(\bo, \phi)
\end{pmatrix},\thinspace
\boldsymbol{{B}}&= 
\begin{pmatrix}
0&\bold c\\
\bold c&0
\end{pmatrix},
\label{A4}
\end{align}
and $
\bold a(\bold k, \phi)= \tilde v_0 \bold{I}_{3\times 3} -2v_D{\bf \Lambda}(\bo,\phi)
$,
\begin{align}
{\bf \Lambda}(\bo,\phi)&=
\begin{pmatrix}
0&\cos k_1 e^{-i\phi}&\cos k_3 e^{i\phi}\\
\cos k_1 e^{i\phi}&0&\cos k_2 e^{-i\phi}\\
\cos k_3 e^{-i\phi}&\cos k_2 e^{i\phi} &0\\ 
\end{pmatrix},
\end{align}
\begin{align}
\boldsymbol{b}=-v_t\cos^2\chi\bold{I}_{3\times 3}, 
\thinspace \bold c =v_t\sin^2\chi\bold{I}_{3\times 3},
\end{align}
where  $\bold{I}_{3\times 3}$ is a $3\times 3$ identity matrix, $\tilde v_0= 4v_J +h\cos\chi - v_t\cos 2\chi= 4v_J +v_t$, $k_i=\bold k\cdot \bold e_i$, $\bold e_1=-(1/2,~\sqrt 3/2),~\bold e_2=(1,0),~\bold e_3=(-1/2,~\sqrt 3/2)$. At the saturation field $h=h_s,~\chi=0$, we obtain ferromagnetically coupled layers applicable to Cu(1-3, bdc)\cite{alex6, alex5a} assuming strong interlayer coupling. 
\begin{figure}
\centering
\includegraphics[width=3in]{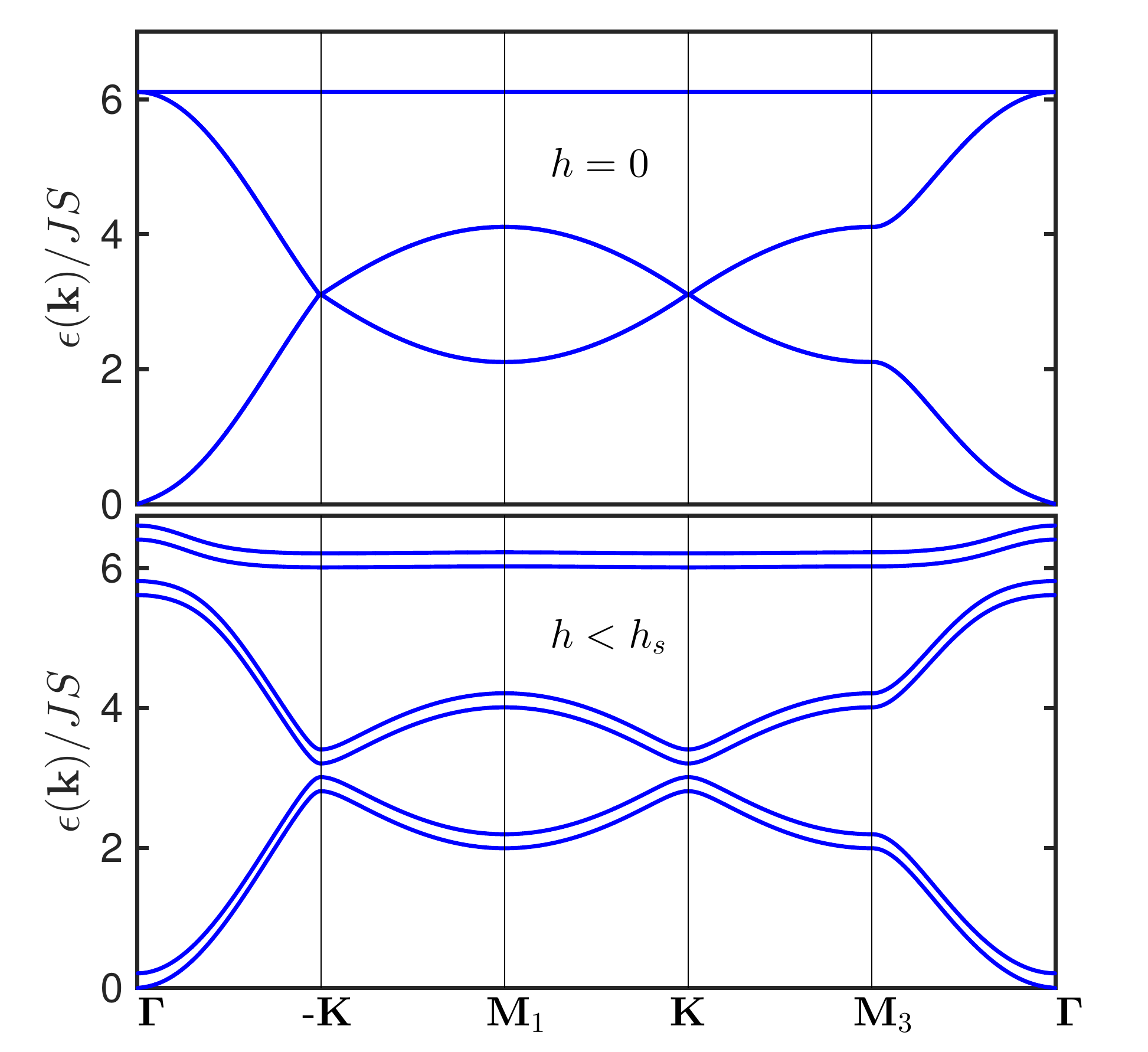}
\caption{Color online. Magnon bands of bilayer Kagom\'e antiferromagnets with $\boldsymbol{\mathcal D}=\mathcal D\bold{\hat z}$ at two values of magnetic field.  $\mathcal D/\mathcal J=0.2,~\mathcal J_t/\mathcal J=0.11$. }
\label{band}
\end{figure}
\begin{figure}
\centering
\includegraphics[width=3in]{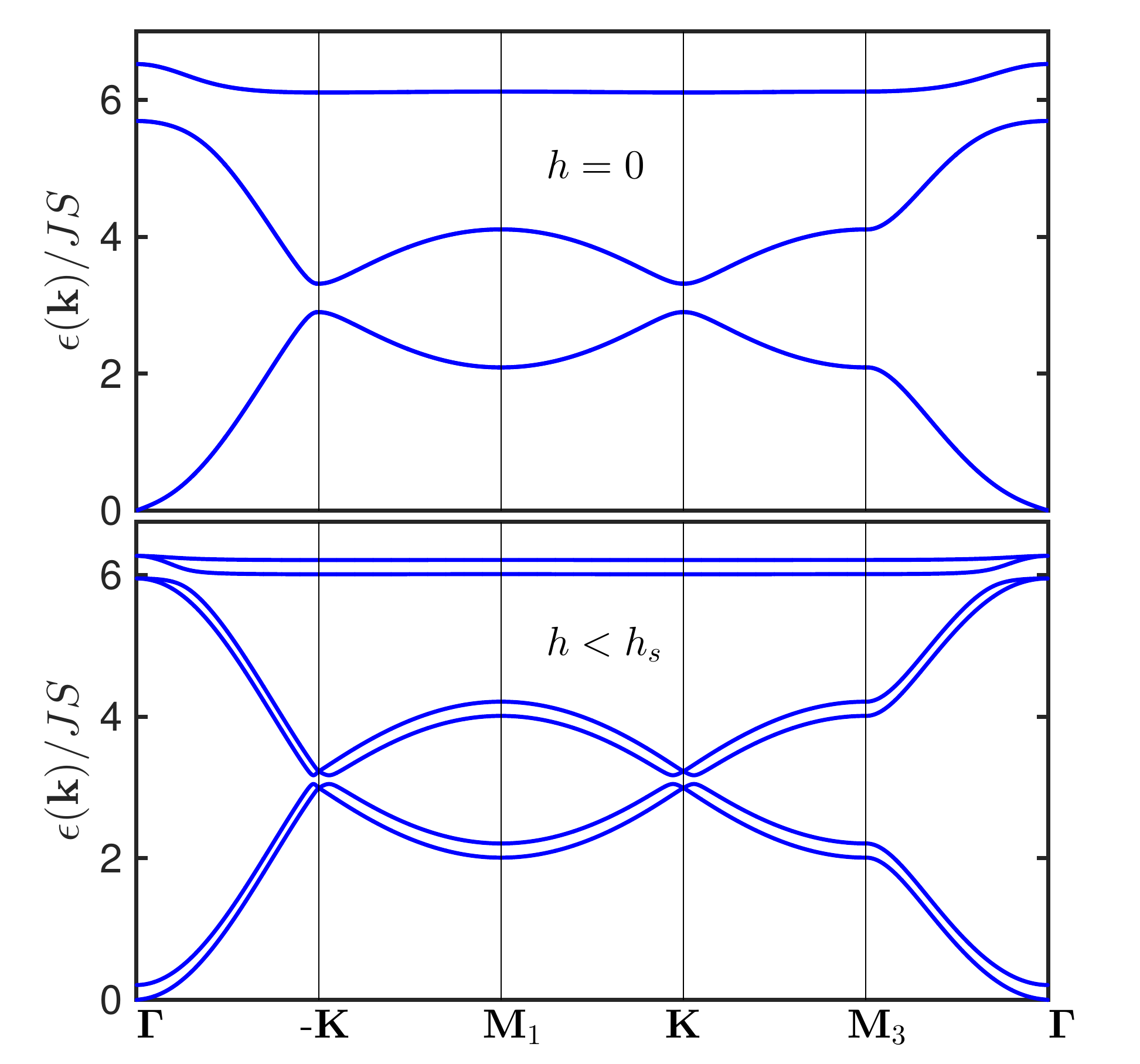}
\caption{Color online. Magnon bands of bilayer Kagom\'e antiferromagnets with $\boldsymbol{\mathcal D}=\mathcal D\bold{\hat x}$ at two values of magnetic field.  $\mathcal D/\mathcal J=0.2,~\mathcal J_t/\mathcal J=0.11$. }
\label{band1}
\end{figure}

\begin{figure}  
\centering
\includegraphics[width=3in]{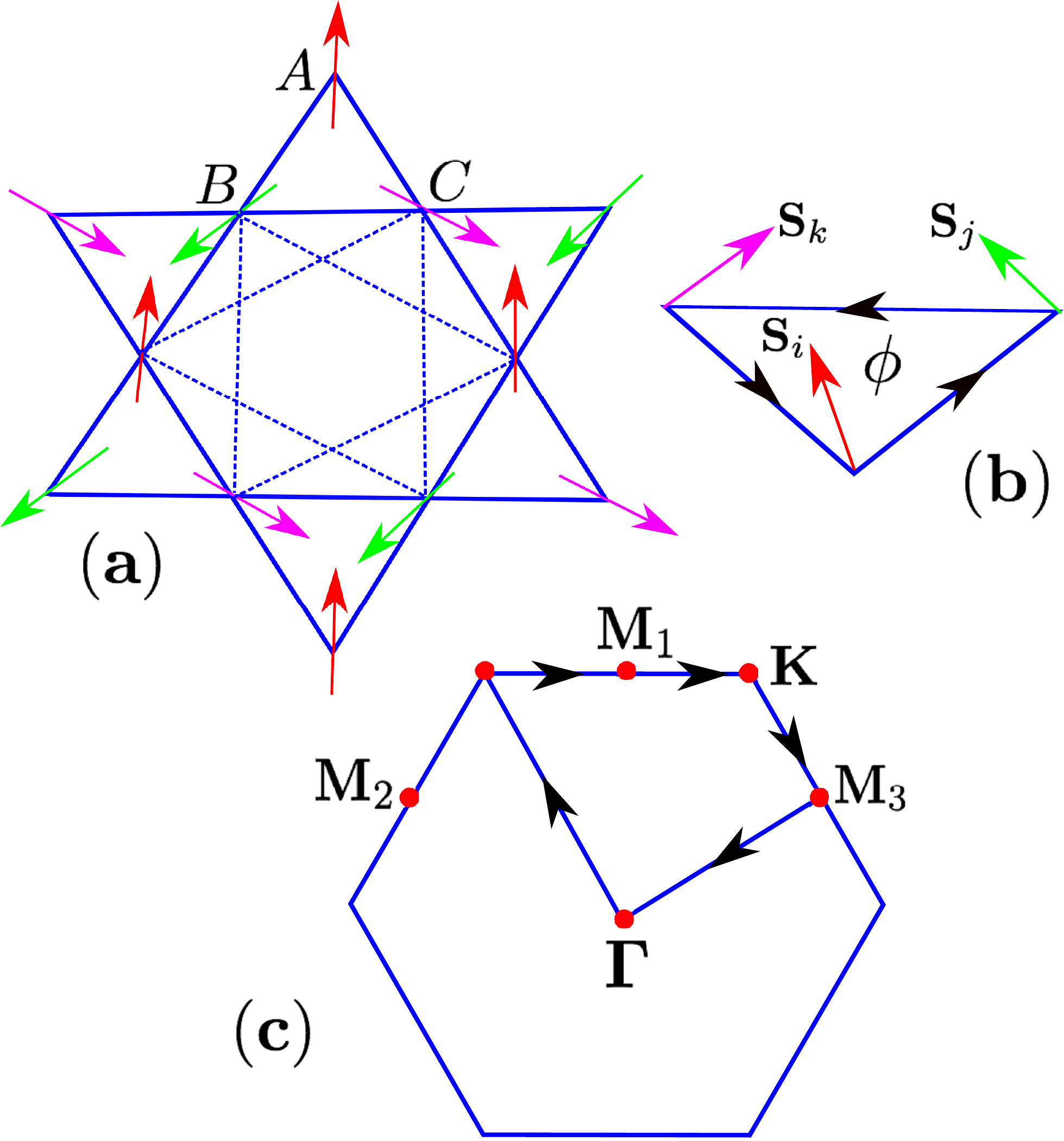}
\caption{Color online. $(a)$~ The zero field  coplanar $120^\circ$ N\'eel order on the  Kagom\'e lattice corresponding to the $\bold q=0$ ground state of Kagom\'e antiferromagnets. Solid lines connect NN sites and dash lines connect NNN sites.  $(b)$~ Field-induced non-coplanar out-of-plane spin canting with nonzero spin scalar chirality $\kappa$, where $\phi$ is the field-induced fictitious flux. }
\label{neel}
\end{figure}

Plotted in Figs.~\ref{band} and \ref{band1} are the magnon bands along the Brillouin zone paths in Fig.~\ref{neel}(c),  with the parameter values of Ca$_{10}$Cr$_7$O$_{28}$, $\mathcal J=0.76~\text{meV},~ \mathcal J_t/\mathcal J=0.11$ \cite{Balz}  and the DM value $\mathcal D/\mathcal J=0.2$. For  $\boldsymbol{\mathcal D}_{ij}=\mathcal D\hat{\bold z}$, the DMI does not contribute at zero field because the spins are along the $x$-$y$ Kagom\'e plane. The resulting magnon bands are doubly degenerate between $S_z \to S_x=\pm S$. At finite  magnetic field increases, each spin has a component along the $z$-axis, hence  the degeneracy of the bands between $S_x=\pm S$ is lifted and the effects of the DMI results in a gap opening at ${\bf K}$.  At  the saturation point $h=h_s$ (not shown) each layer is fully polarized along the field $z$-direction, again the DMI leads to gapped non-degenerate magnon bands. For the in-plane DMI $\boldsymbol{\mathcal D}_{ij}=\mathcal D\hat{\bold x}$ the situation is different.  The degeneracy persists at zero field but since the spins are along the $xy$ plane the DMI leads to  gap magnon bands. The degeneracy is always lifted at nonzero magnetic field, but in this case the there is a staggered flux emanating from both layers and the bands cross each other at $\pm {\bf K}$.  At the saturation point $h=h_s$ the spin are fully polarized along the $z$-axis and the in-plane DMI does not contribute and the non-degenerate magnon bands are  gapless  (not shown). 

\subsection{Model II}
In the previous section, we studied one of the possible field-induced magnetically ordered phases in geometrically frustrated bilayer Kagom\'e magnets.  In this section, we study another magnetically ordered phase which has been realized experimentally in various frustrated  Kagom\'e magnets. Without loss of generality, we focus on the  ideal Kagom\'e material KFe$_3$(OH)$_{6}$(SO$_{4}$)$_2$ \cite{sup1a,sup2}. The Hamiltonian is given by  \begin{align}
\mathcal H&= \sum_{ i, j} \mathcal J_{ij}{\bf S}_{i}\cdot{\bf S}_{j} + \sum_{\la i, j\ra} \boldsymbol{\mathcal D}_{ij}\cdot{\bf S}_{i}\times{\bf S}_{j}-h\hat{\bold z}\cdot\sum_i \bold{S}_{i}
\label{apen1},
\end{align}
where $\mathcal J_{ij}=\mathcal J>0 $ and $\mathcal J_2>0$ are the isotropic antiferromagnetic couplings for nearest-neighbour (NN) and  next-nearest-neighbour (NNN) sites respectively,   $\boldsymbol{\mathcal D}_{ij}=(0,0, \mp \mathcal D_z)$, where $-/+$ denotes the directions of the out-of-plane DMI in the up/down triangles of the Kagom\'e lattice, and $h$ is the out-of-plane magnetic field in units of $g\mu_B$. For the iron jarosite KFe$_3$(OH)$_{6}$(SO$_{4}$)$_2$ the interlayer exchange interaction  can be neglected for two reasons. First, it is very small compare to $\mathcal J $ and $\mathcal J_2$. Second, a single crystal of iron jarosite can be synthesized \cite{sup2}.   We have also neglected the in-plane DM vector because it neither stabilizes magnetic ordering nor induces topological effects. Therefore, it  cannot change  the results obtain here.
\begin{figure}
\centering
\includegraphics[width=3in]{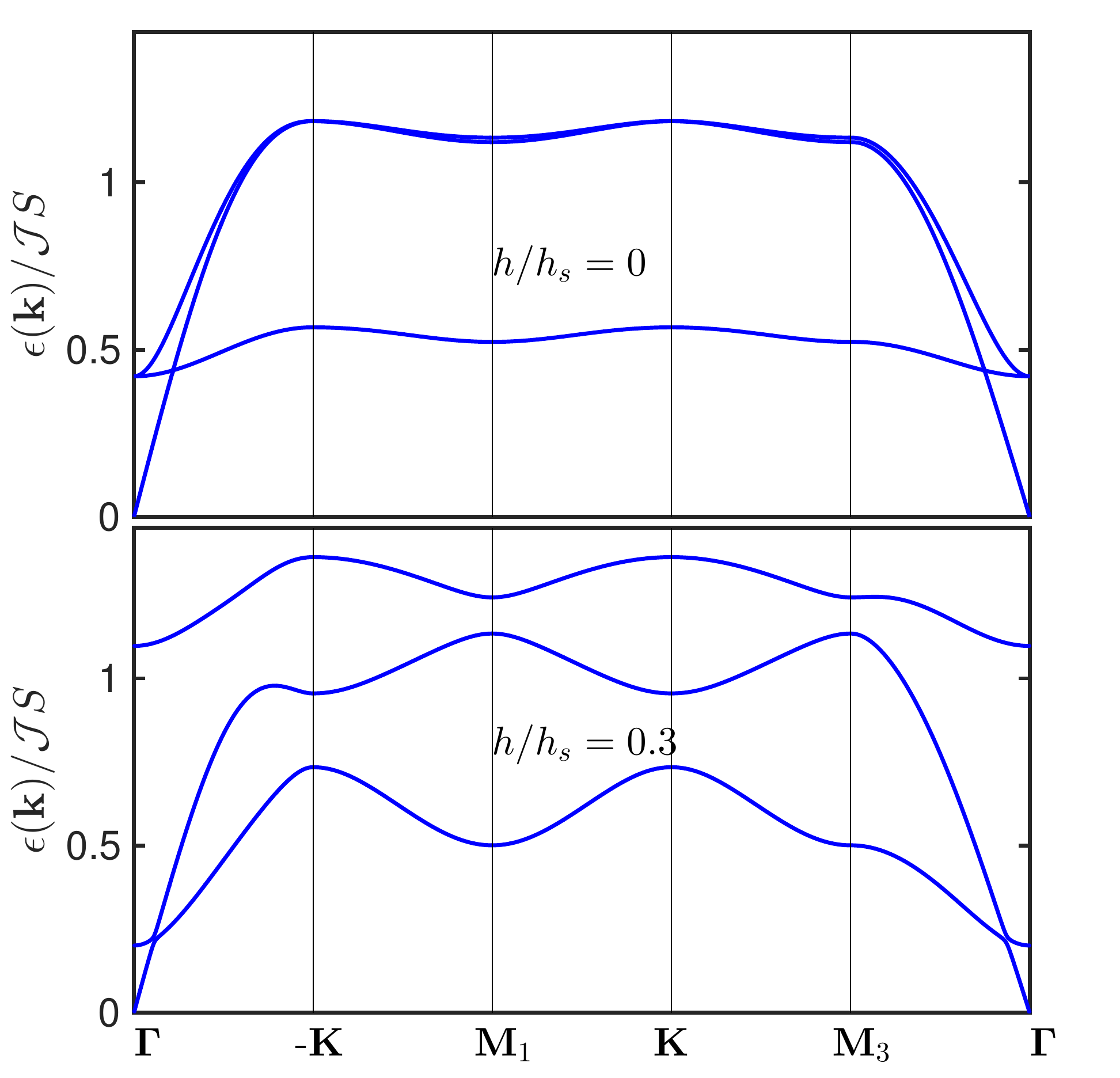}
\caption{Color online. Magnon bands of iron jarosite KFe$_3$(OH)$_{6}$(SO$_{4}$)$_2$ with $\boldsymbol{\mathcal D}=\mathcal D\bold{\hat z}$ at two values of magnetic field.  $\mathcal D/\mathcal J=0.06,~\mathcal J_2/\mathcal J=0.03$.}
\label{band2}
\end{figure}
\begin{figure}
\centering
\includegraphics[width=3in]{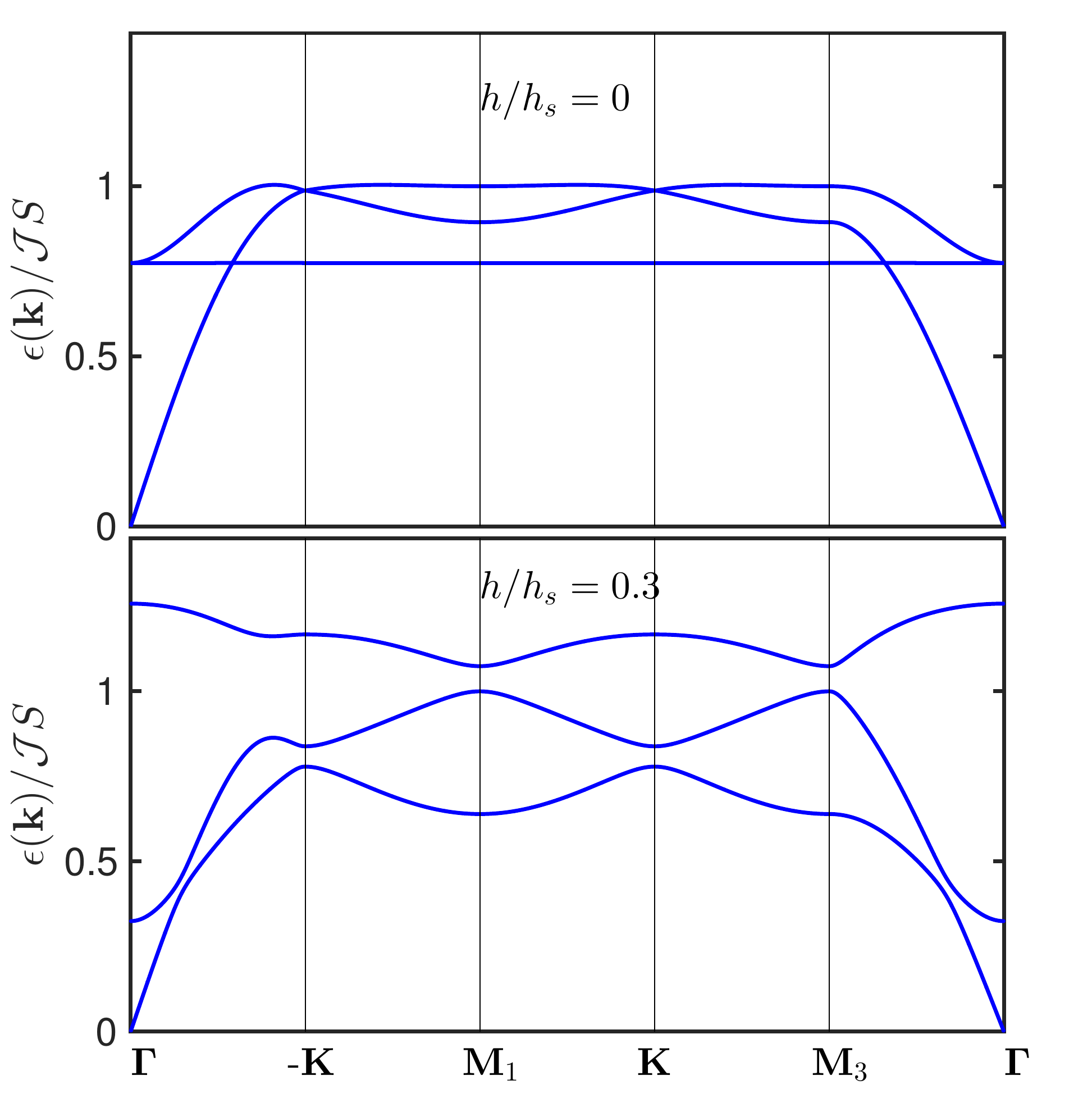}
\caption{Color online. Magnon bands of XXZ Kagom\'e antiferromagnets without DMI at two values of magnetic field.  $\delta=0.4$.}
\label{band3}
\end{figure}
The  alternating out-of-plane DMI between the up and down triangles of the Kagom\'e lattice is fictitious in that  only one ground state is selected for each sign with positive chirality $(\mathcal D_z >0)$ or negative chirality $(\mathcal D_z < 0)$.  At zero magnetic field, the out-of-plane DMI induces a coplanar 120$^\circ$ N\'eel order on the $x$-$y$ Kagom\'e plane with positive chirality \cite{sup1,sup1a, sup2} as shown in Fig.~\ref{neel}(a). For nonzero out-of-plane field, inelastic neutron scattering experiment on the ideal Kagom\'e material KFe$_3$(OH)$_{6}$(SO$_{4}$)$_2$ \cite{sup1,sup1a,sup2} has uncovered a non-coplanar spin texture with nonzero spin chirality $\kappa=\sum{\bf S}_i\cdot\lb{\bf S}_j\times {\bf S}_k\rb$ \cite{sup1a} as shown in Fig.~\ref{neel}(b). However, the topological effects of this non-collinear system have not been studied both theoretically and experimentally. In a recent Letter \cite{me}, we showed for the first time that this material possesses a finite thermal Hall conductivity in this non-collinear regime. Thus, the iron jarosite  KFe$_3$(OH)$_{6}$(SO$_{4}$)$_2$ is a possible candidate for investigating thermal Hall effect of magnon, which is accessible by using inelastic neutron scattering.


 In the present study, we analyze the finite thermal Hall conductivity in terms of topological magnon edge states.   We consider the $\bold q=0$ ground state with  positive chirality  $\boldsymbol{\mathcal D}_{ij}=(0,0, -\mathcal D_z)$ and $\mathcal D_z>0$. The basic  procedure is similar to the bilayer system studied above. The rotation matrix is same as Eq.~\ref{rot}, however, with  the oriented angles $\theta_A=0,~\theta_B=2\pi/3,~\theta_C=-2\pi/3$.  The classical energy is given by \begin{align}
e_{cl}= \tilde {\mathcal J}\lb -1 + 3\cos^2\chi\rb -\sqrt{3}\mathcal D_z\sin^2\chi-h\cos\chi,
\end{align}
where $e_{cl}=E_{cl}/3NS^2$, $\tilde {\mathcal J}=\mathcal J +\mathcal J_2$, and the magnetic field is rescaled in units of $S$. Minimizing $e_{cl}$ yields the canting angle $\cos\chi = h/h_s$, with $h_s=(6\tilde {\mathcal J}+2\sqrt{3}\mathcal D_z)$. We see that the DMI depends on the classical energy as it contributes to the stability of the $\bold q=0$ ground state. Besides, $\mathcal J_2>0$ can also stabilize the coplanar magnetic structure in the absence of the DMI.   The topological excitations above the classical ground state is mediated by a field-induced scalar chirality $\kappa_{\chi}=\cos\chi\sum{\bf S}_i\cdot\lb{\bf S}_j\times {\bf S}_k\rb$ defined as the solid angle subtended by three neighbouring spins. As mentioned previously, the scalar spin chirality originates from non-coplanar spin texture and does not need the presence of DMI or magnetic ordering. It is also the basis of chiral spin liquid states which suggests that topological effects may persist in the spin liquid regime of frustrated kagom\'e magnets.  This model does not have an analogy to collinear ferromagnets \cite{alex1, alex0, alex2,alex5,alex4, sol1,sol,sol2, alex5a, alex6,    mok,su,fyl,lifa} or bilayer collinear in Model I.  It also differs significantly from field-induced topological magnons in bipartite frustrated honeycomb lattice because $\kappa_{\chi}=0$ \cite{sol4}. In fact, the bipartite honeycomb antiferromagnets fall into the class of Model I as they are doubly degenerate at zero magnetic field and require an explicit DMI.  The magnon tight binding Hamiltonian for Model II is given by
 \begin{align}
&\mathcal H_{SW}= S\sum_{\bo,\alpha,\beta; 1,2}2\lb \mathcal{M}_{\alpha\beta}^0\delta_{\alpha\beta} +\mathcal{M}_{\alpha\beta; 1,2}\rb b_{\bo \alpha}^\dagger b_{\bo \beta}\label{main}\\&\nonumber +\mathcal{M}_{\alpha\beta; 1,2}^{\prime} \lb b_{\bo \alpha}^\dagger b_{-\bo \beta}^\dagger +b_{\bo \alpha} b_{-\bo \beta}\rb,
\end{align}
where $\alpha,\beta=A,B,C$ and the coefficients are given by
$
\boldsymbol{\mathcal{M}_0}=\zeta\bold{I}_{3\times 3},$ with $\zeta=( \tilde {\mathcal J} +\sqrt{3}\mathcal D_z)$.

\begin{align}
&\boldsymbol{\mathcal{M}}_{1,2}= \Delta_{1,2}\begin{pmatrix}
  0& \gamma_{AB}^{1,2}e^{-i\phi_{1,2}}&\gamma_{CA}^{1,2}e^{i\phi_{1,2}} \\
\gamma_{AB}^{*1,2} e^{i\phi_{1,2}}& 0&\gamma_{BC}^{1,2}e^{-i\phi_{1,2}}\\
\gamma_{CA}^{*1,2} e^{-i\phi_{1,2}}& \gamma_{BC}^{*1,2} e^{i\phi_{1,2}} & 0 \\  
 \end{pmatrix};\\
 &\boldsymbol{\mathcal{M}}_{1,2}^\prime=\Delta_{1,2}^\prime\begin{pmatrix}
  0& \gamma_{AB}^{1,2}&\gamma_{CA}^{1,2} \\
\gamma_{AB}^{*1,2} & 0&\gamma_{BC}^{1,2}\\
\gamma_{CA}^{*1,2} & \gamma_{BC}^{*1,2} & 0 \\  
 \end{pmatrix};
\end{align}
where $\gamma_{AB}^1=\cos k_1,~ ~\gamma_{BC}^1=\cos k_2,  \gamma_{CA}^1=\cos k_3;~\gamma_{AB}^2=\cos p_1,~ ~\gamma_{BC}^2=\cos p_2,  \gamma_{CA}^2=\cos p_3$ and $p_i=\bold{p}\cdot\bold{e}_i^\prime,~\bold e_1^\prime=\bold e_3-\bold e_2,~\bold e_2^\prime=\bold e_1-\bold e_3,~\bold e_3^\prime=\bold e_2-\bold e_1$. The fictitious magnetic fluxes are given by  $\phi_{1,2}=\tan^{-1}\lb\Delta_{1,2}^M/\Delta_{1,2}^R\rb$, and $\Delta_{1,2}=\sqrt{(\Delta_{1,2}^R)^2+(\Delta_{1,2}^M)^2}$, where 
\begin{align}
  &\Delta^R_{1}= \mathcal J\lb -\frac{1}{2} +\frac{3}{4}\sin^2\chi\rb-\frac{\sqrt 3 \mathcal D_z}{2}\lb 1-\frac{\sin^2\chi}{2}\rb,\\&\Delta_1^M=\frac{\cos\chi}{2}\lb -\sqrt{3}\mathcal J +\mathcal D_z\rb,\\& \Delta_{1}^{\prime}=\frac{\sin^2\chi}{4}\lb 3\mathcal J+\sqrt 3\mathcal D_z\rb, \\&\Delta^{R(M)}_{2}=\Delta^{R(M)}_{1}(\mathcal D_z\to 0, \mathcal J\to \mathcal J_2),\\&
  \Delta_2^\prime= \Delta_1^\prime (\mathcal D_z\to 0, \mathcal J\to \mathcal J_2).
 \end{align}
Using the vector notation \bea \Psi^\dg_\bo= (b_{\bo A}^{\dg},\thinspace b_{\bo B}^{\dg},\thinspace b_{\bo C}^{\dg}, \thinspace b_{-\bo A},\thinspace b_{-\bo B},\thinspace b_{-\bo C} ),\eea the Hamiltonian can be written as \bea
\mathcal H_{SW}=\mathcal{E}_0+ S\sum_{\bo} \Psi^\dg_\bo \boldsymbol{ \mathcal{H}}(\bo)\Psi_\bo, \eea 
where  
\begin{align}
\boldsymbol{\mathcal{H}}(\bo)=\mathbb{I}_{2\times 2}\otimes\lb\boldsymbol{\mathcal M_0}+\boldsymbol{\mathcal M}\rb +\sigma_x\otimes \boldsymbol{\mathcal{M}}^\prime, 
\end{align}
and $\mathcal{E}_0$ is a constant.  $\mathbb{I}_{2\time 2}$ is an identity $2\times 2$ matrix and $\sigma_x$ is a Pauli matrix. $\boldsymbol{\mathcal M}=\boldsymbol{\mathcal M}_1+\boldsymbol{\mathcal M}_2$, the same for $\boldsymbol{\mathcal M}^\prime$. The eigenvalues of this Hamiltonian cannot be obtained  analytically as opposed to the zero field case, $\chi=\pi/2$, with coplanar 120$^\circ$ N\'eel order on the $x$-$y$ Kagom\'e plane. It is important to note that both fluxes do not vanish at zero DMI. This means that the DMI does not provide topological effects  in stark contrast to ferromagnets \cite{alex1, alex0, alex2,alex5,alex4, sol1,sol,sol2, alex5a, alex6,  su, lifa,mok}. As shown in Fig.~\ref{band2} the magnon bands are not topological at zero magnetic field even in the presence of DMI.


\subsection{Model III}
The final model we shall  consider is the XXZ Kagom\'e antiferromagnet without DMI subject to an out-of-plane magnetic field.   The  Hamiltonian is governed by 
\begin{align}
\mathcal H&= \mathcal J\sum_{ \la i, j\ra}\lb {\bf S}_{i}\cdot{\bf S}_{j}-\delta S_{i}^zS_{j}^z\rb -h\hat{\bold z}\cdot\sum_i \bold{S}_{i}\label{eq1},
\end{align}
where $\mathcal J>0$  and $0\leq \delta\leq 1$ is the easy-plane anisotropy, and $h$ is the magnetic field in units of $g\mu_B$. At zero field, the easy-plane anisotropy  favours the positive chirality $\bold q=0$ ground state \cite{sup1,sup1a, sup2,sup3}. The canting angle is determined from the classical energy
\begin{align}
e_{cl}= \mathcal J[ -1+  \lb 3-2\delta\rb\cos^2\chi]-h\cos\chi,
\end{align}
where $\cos\chi = h/h_s$ and $h_s=2J(3-2\delta)$. 

In this system, topological magnon bands originate from $\kappa_{\chi}$ as in Model II.  The fictitious magnetic flux encountered by propagating magnon is given by $\tan\phi=\Delta_M/\Delta_R$ where $\Delta=\sqrt{\Delta_R^2+\Delta_M^2}$, and 
\begin{align}
  &\Delta_R= \bigg[ -\frac{1}{2} +\frac{1}{2}\lb\frac{3}{2}-\delta\rb\sin^2\chi\bigg],\\&\Delta_M=-\frac{\sqrt{3} }{2}\cos\chi,~
  \Delta^\prime= \frac{\sin^2\chi}{2}\lb\frac{3}{2}-\delta\rb.
 \end{align}
In Fig.~\ref{band3} we show the magnon bands for $\delta=0.4$. At zero field, the magnon bands are very similar to Model II.  We see that the  flat mode  acquires a small dispersion at nonzero field and the magnon bands are gapped at all points in the Brillouin zone.  

\subsection{Finite thermal Hall conductivity at zero DMI}

\begin{figure}[!]
\centering
\includegraphics[width=3.75in]{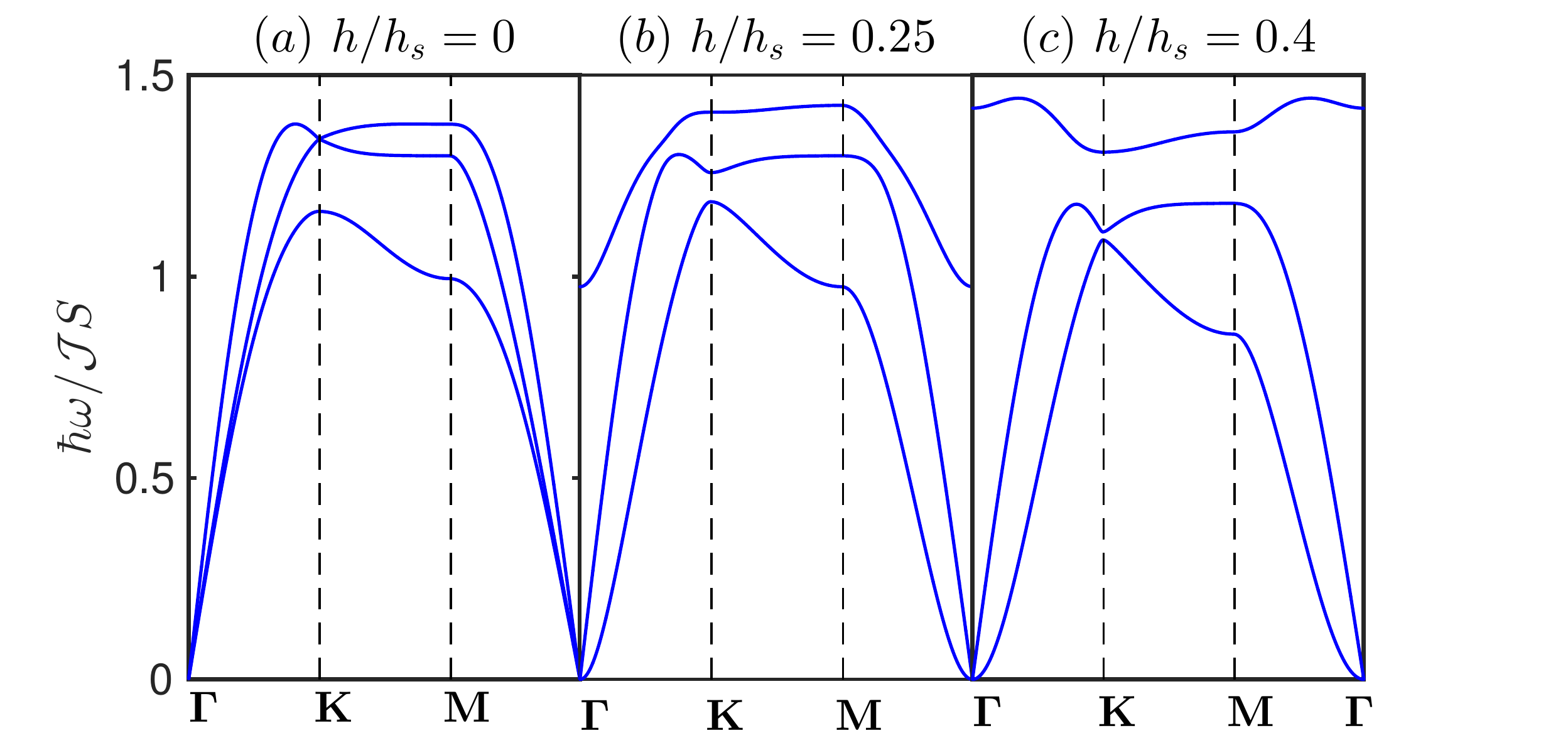}
\caption{Color online.  Magnon bands of noncollinear $\bold q=0$ kagom\'e antiferromagnet with zero DMI  at three field values. The parameters are $\mathcal D_z/\mathcal J=0.0,~\mathcal J_2/\mathcal J=0.3$.}
\label{band4}
\end{figure}

\begin{figure}[!]
\centering
\includegraphics[width=3in]{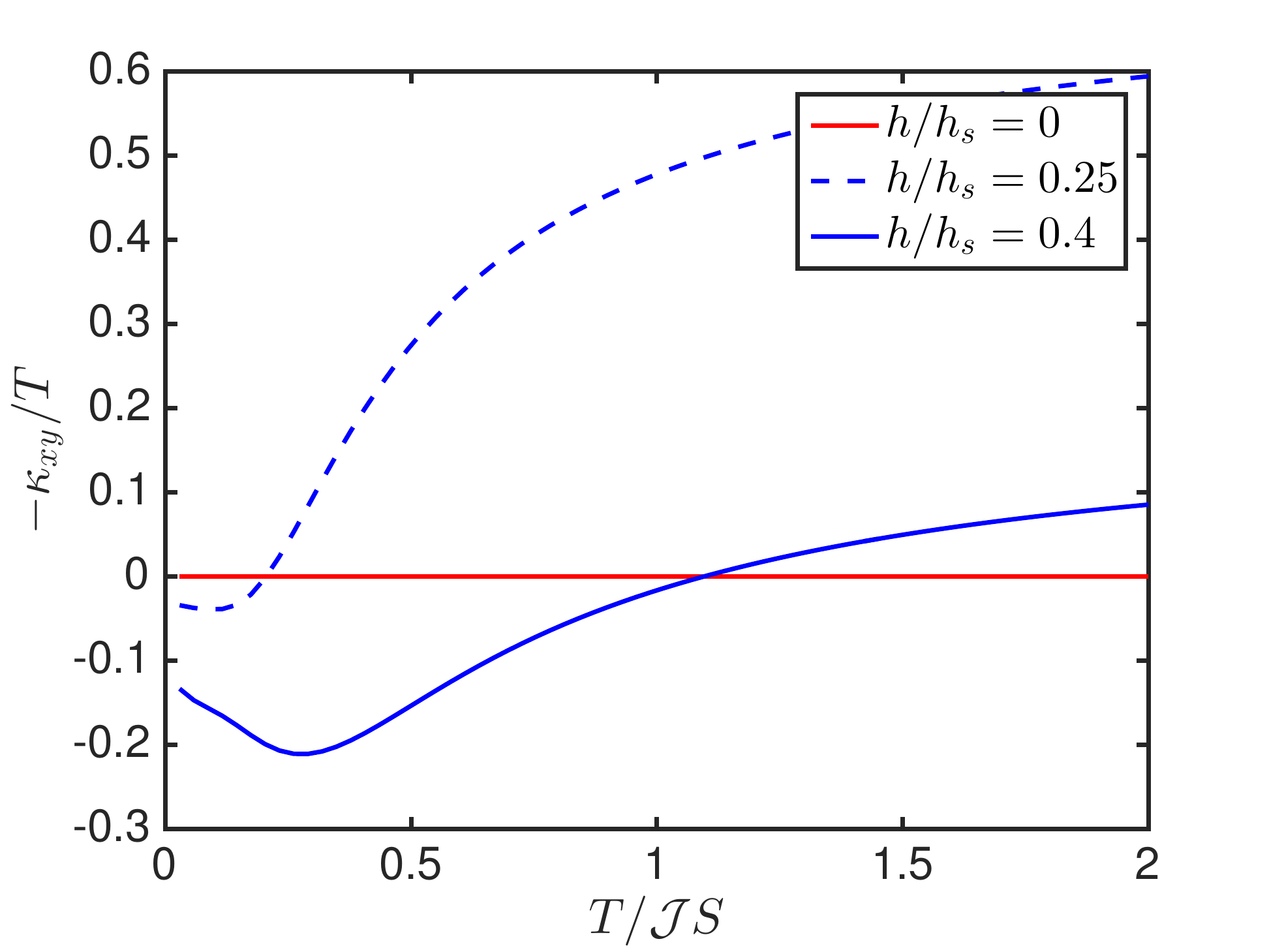}
\caption{Color online.  Low-temperature dependence of $\kappa_{xy}$  for the bands in Fig.~\ref{band4} at three field values.}
\label{N_thd}
\end{figure}

As mentioned in the text, the DMI does not provide any topological effects for the noncollinear $\bold q=0$ spin configuration on the kagom\'e lattice. This means that topological effects persist at zero DMI and nonzero out-of-plane magnetic field via an induced spin scalar chirality  $\mathcal H_{\chi}\sim\cos\chi \sum {\bold S}_i\cdot\lb \bold S_{j}\times\bold S_{k}\rb$, where $\cos\chi=h/h_s$ with $h_s= 6(\mathcal J+\mathcal J_2)$. However,   DMI is usually present on the kagom\'e lattice due to lack of inversion center.  In this section, we show that anomalous magnon Hall effect is present at zero DMI in contrast to collinear ferromagnets.  Figures~\ref{band4} and \ref{N_thd} show the magnon bands and the corresponding $\kappa_{xy}$ respectively for zero DMI.

\section{Topological Magnon Edge Modes}
The Hamiltonians for insulating antiferromagnets are diagonalized by the generalized Bogoliubov transformation 
$\Psi_\bo= \mathcal{P}_\bo Q_\bo$, 
where $\mathcal{P}_\bo$ is a $2N\times 2N$ paraunitary matrix and  $Q^\dg_\bo= (\mathcal{Q}_\bo^\dg,\thinspace \mathcal{Q}_{-\bo})$ with $ \mathcal{Q}_\bo^\dg=(\beta_{\bo A}^{\dg}\thinspace \beta_{\bo B}^{\dg}\thinspace \beta_{\bo C}^{\dg})$ being the quasiparticle operators. The matrix $\mathcal{P}_\bo$ satisfies the relations,
\begin{align}
&\mathcal{P}_\bo^\dg \boldsymbol{\mathcal{H}}(\bo) \mathcal{P}_\bo=\mathcal{E}_\bo\label{eqn1}\\ &\mathcal{P}_\bo^\dg \boldsymbol{\tau}_3 \mathcal{P}_\bo= \boldsymbol{\tau}_3,
\label{eqna}
\end{align}
where $\mathcal{E}_\bo= \text{diag}(\epsilon_{\bo\alpha},~\epsilon_{-\bo\alpha})$, 
$ \boldsymbol{\tau}_3=
\text{diag}(
 \mathbf{I}_{N\times N}, -\mathbf{I}_{N\times N} )$, and $\epsilon_{\bo\alpha}$ are the  energy eigenvalues.
From Eq.~\ref{eqna} we get $\mathcal{P}_\bo^\dg= \boldsymbol{\tau}_3 \mathcal{P}_\bo^{-1} \boldsymbol{\tau}_3$, and Eq.~\ref{eqn1} is equivalent to saying that we need to diagonalize the Hamiltonian $\boldsymbol{\mathcal{H}}^\prime(\bo)= \boldsymbol{\tau}_3\boldsymbol{\mathcal{H}}(\bo),$
 whose eigenvalues are given by $ \boldsymbol{\tau}_3\mathcal E_\bo$ and the columns of $\mathcal P_\bo$ are the corresponding eigenvectors.  The paraunitary operator defines a Berry curvature given by
 \begin{align}
 \Omega_{ij;\alpha}(\bo)=-2\text{Im}[ \boldsymbol{\tau}_3\mathcal (\partial_{k_i}\mathcal P_{\bo\alpha}^\dg) \boldsymbol{\tau}_3(\partial_{k_j}\mathcal P_{\bo\alpha})]_{\alpha\alpha},
 \label{bc1}
 \end{align}
with $i,j=\lbrace x,y\rbrace$ and $\mathcal P_{\bo\alpha}$ are the columns of $\mathcal P_{\bo}$. This form of the Berry curvature simply extracts the diagonal components which are the most important. Suppose the explicit analytical form of $\mathcal P_{\bo\alpha}$ is known as in honeycomb-lattice hardcore bosons \cite{sol2}, the Berry curvature can be computed directly from Eq.~\ref{bc1}.  Unfortunately, this is not the case in the present models. From Eq.~\ref{eqn1} the Berry curvature can be written alternatively as

\begin{align}
\Omega_{ij;\alpha}(\bold k)=-2\sum_{\alpha^\prime\neq \alpha}\frac{\text{Im}[ \braket{\mathcal{P}_{\bo\alpha}|v_i|\mathcal{P}_{\bo\alpha^\prime}}\braket{\mathcal{P}_{\bo\alpha^\prime}|v_j|\mathcal{P}_{\bo\alpha}}]}{\lb\epsilon_{\bo\alpha}-\epsilon_{\bo\alpha^\prime}\rb^2},
\label{chern2}
\end{align}
where $\bold v=\partial\boldsymbol{\mathcal{H}}^\prime(\bo)/\partial \bold k$ defines the velocity operators. The present form can be computed once the eigenvalues and eigenvectors of the Hamiltonian are obtained numerically. Similar to fermionic systems, the Chern number can still be defined for bosonic systems as the integration of the Berry curvature over the first Brillouin zone,
 \begin{equation}
\mathcal{C}_\alpha= \frac{1}{2\pi}\int_{{BZ}} dk_idk_j~ \Omega_{ij;\alpha}(\bold k).
\label{chenn}
\end{equation}  
Indeed, topologically protected  magnon edge states are characterized by nonzero Chern numbers. However, the Chern numbers are fictitious because the notion of completely filled bands and Fermi energy do not apply to bosonic (magnonic) systems.

\begin{figure}
\centering
\includegraphics[width=3in]{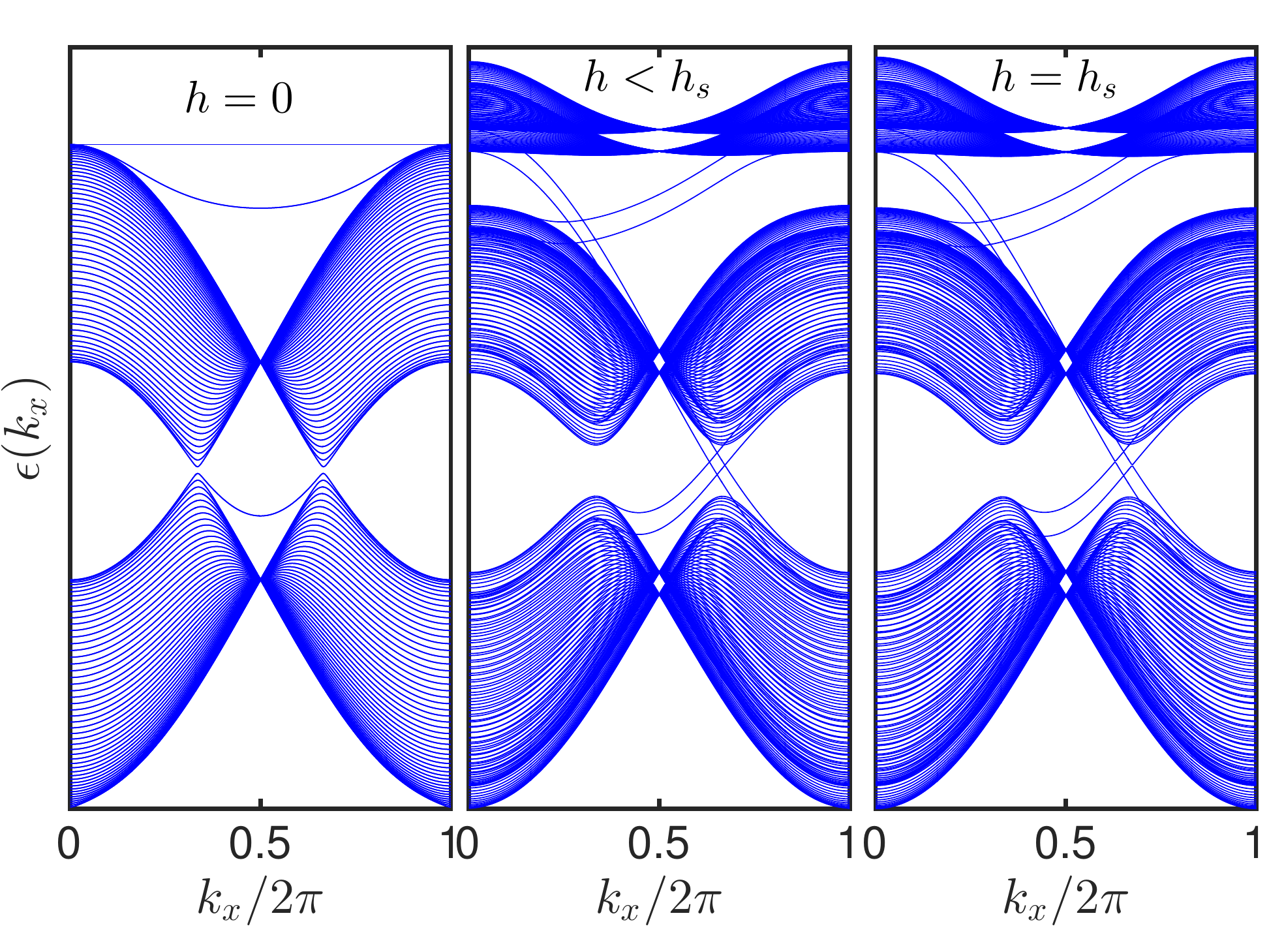}
\caption{Color online. Magnon edge states of bilayer Kagom\'e antiferromagnets (Model I) for a strip geometry  with $\boldsymbol{\mathcal D}_{ij}=\mathcal D\hat{\bold z}$. The parameters are $\mathcal D/\mathcal J=0.2,~\mathcal J_t/\mathcal J=0.11$.}
\label{mo1}
\end{figure}

\begin{figure}
\centering
\includegraphics[width=3in]{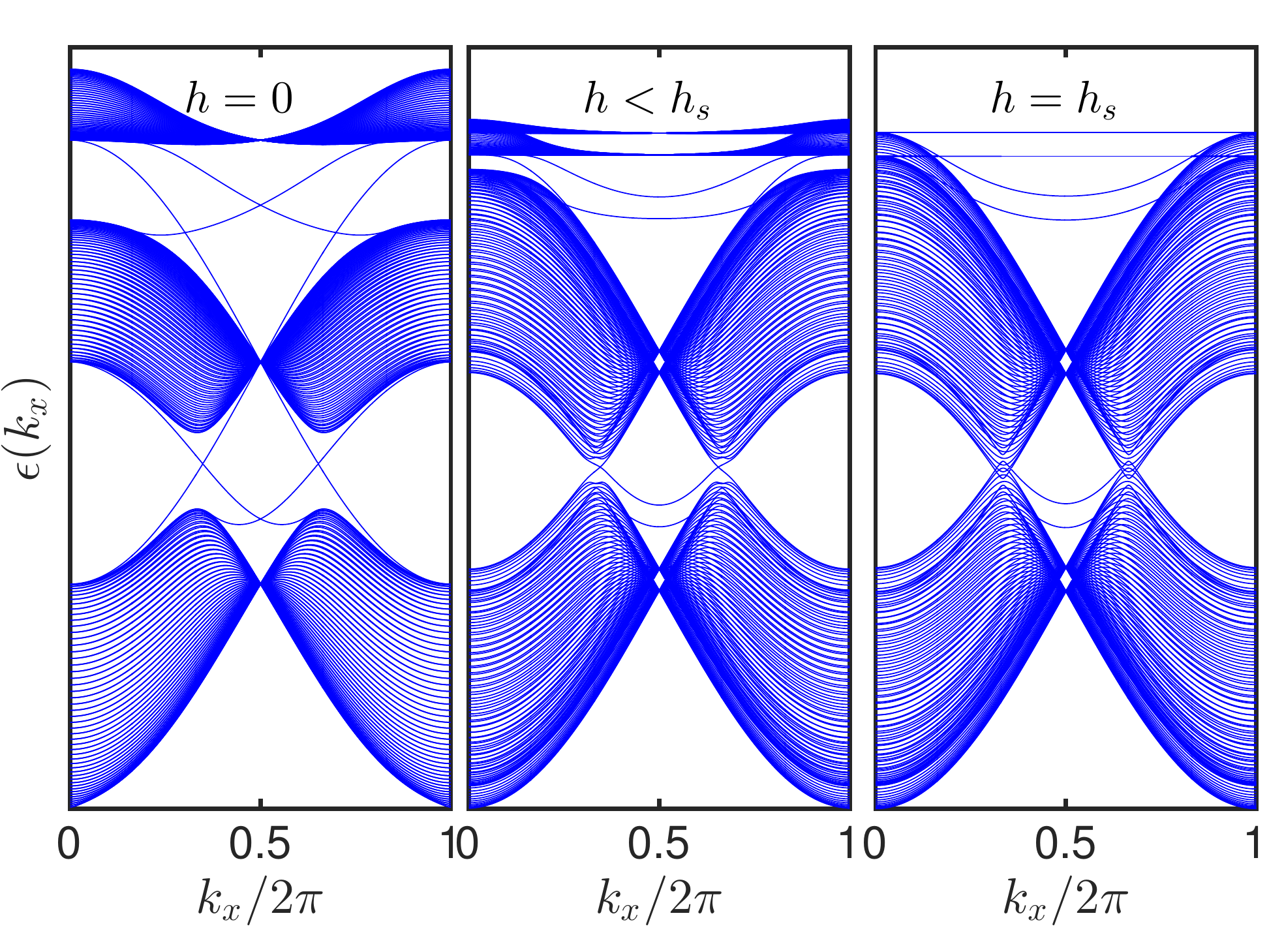}
\caption{Color online. Magnon edge states of bilayer Kagom\'e antiferromagnets (Model I) for a strip geometry with $\boldsymbol{\mathcal D}_{ij}=\mathcal D\hat{\bold x}$. The parameters are $\mathcal D/\mathcal J=0.2,~\mathcal J_t/\mathcal J=0.11$.}
\label{mo2}
\end{figure}

 \begin{figure}
\centering
\includegraphics[width=1\linewidth]{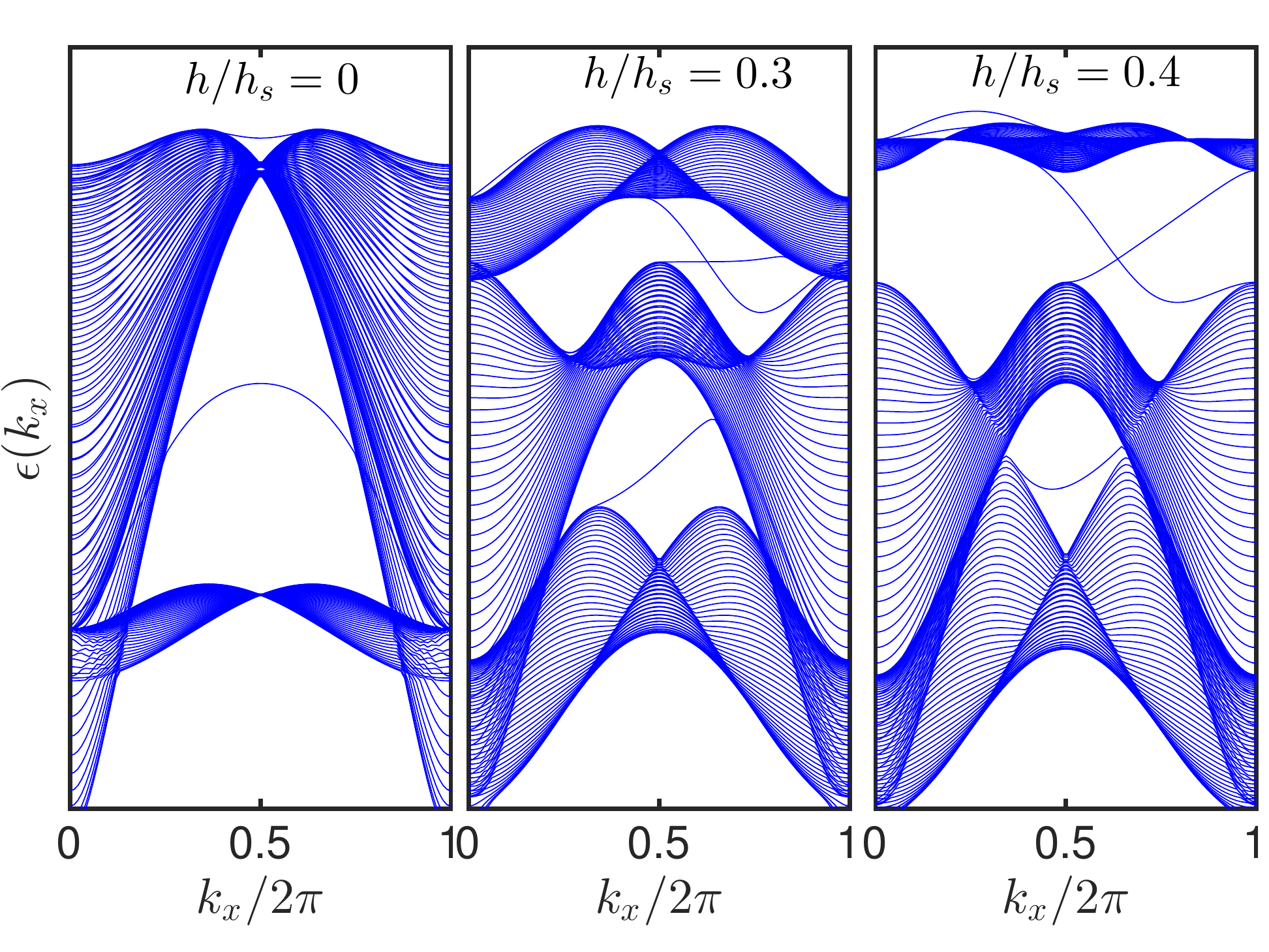}
\caption{Color online.  Magnon edge states of  Kagom\'e antiferromagnets for Model II with the  parameter values of jarosite  KFe$_3$(OH)$_{6}$(SO$_{4}$)$_2$ $\mathcal D_z/\mathcal J=0.06,~\mathcal J_2/\mathcal J=0.03$. }
\label{mo3}
\end{figure}

 \begin{figure}
\centering
\includegraphics[width=1\linewidth]{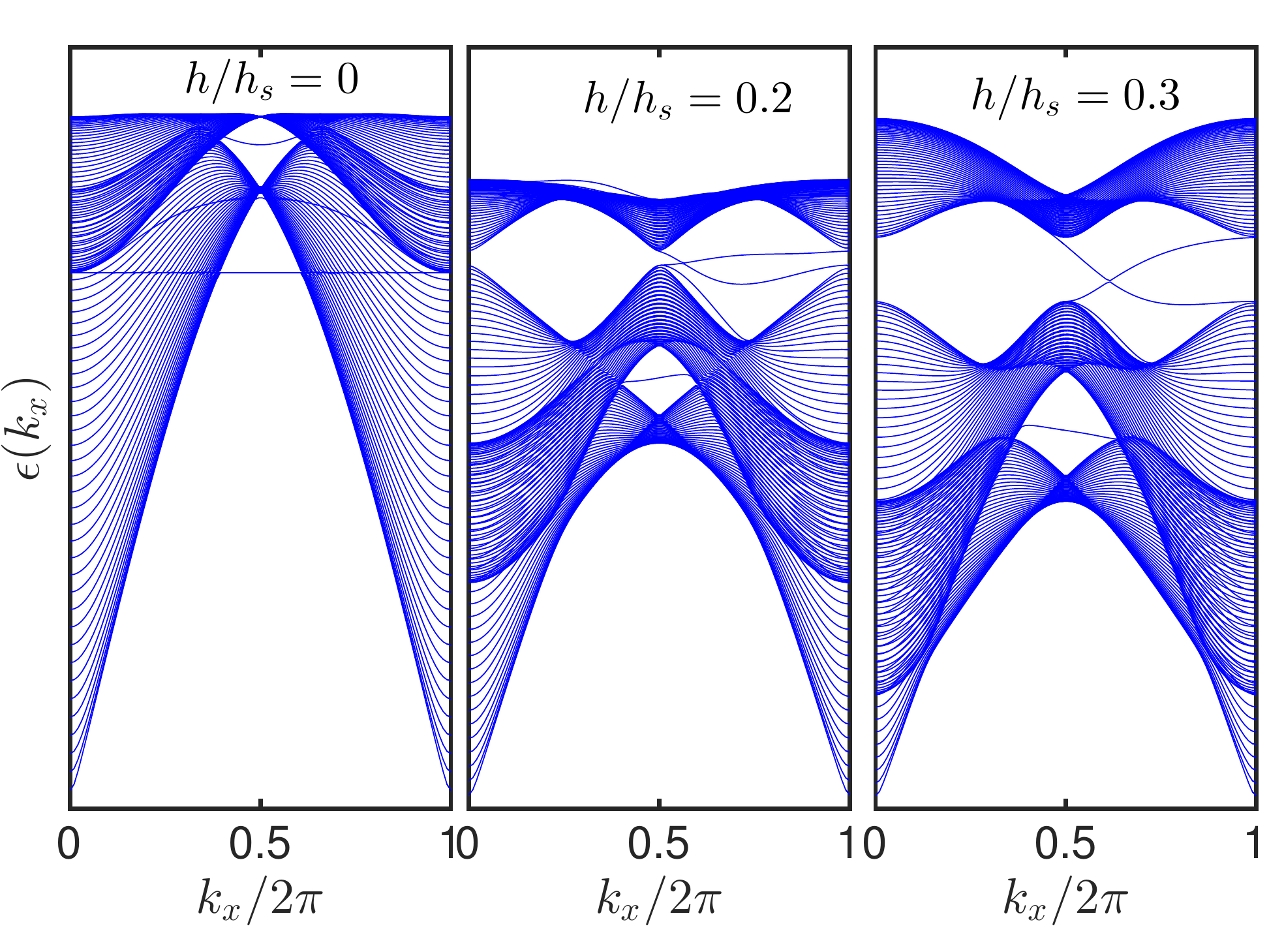}
\caption{Color online.  Magnon edge states of  Kagom\'e antiferromagnets for Model III  with $\delta=0.4$. }
\label{mo4}
\end{figure}
 
Experimentally accessible magnon Hall effect has been previously analyzed in Ref.~\cite{me}.  Now, we complete this study by providing evidence of topological magnon insulator with Chern number protected magnon edge modes. We have solved  for the magnon edge modes using a strip geometry with open boundary conditions along the $y$ direction  and infinite along $x$ direction \cite{guo}.  First, let us consider Model I with $\boldsymbol{\mathcal D}_{ij}=\mathcal D\hat{\bold z}$  shown in Fig.~\ref{mo1}.  In this case the magnon bulk bands are degenerate between $S_z\to S_x=\pm S$ sectors at $h=0~ (\chi=\pi/2)$ and the DMI vanishes in the noninteracting limit as the spins are along the $x$-$y$ Kagom\'e plane at zero field. This results in gapless magnon bulk bands at $\pm{\bf K}$ and ${\bf \Gamma}$ with a single  edge mode connecting these points and  $\kappa_{xy}$ vanishes as expected. For $h<h_s$ the spins are non-collinear and the degeneracy between $S_z\to S_x=\pm S$ is lifted because each spin has a component along the $z$-axis. The DMI opens a small gap between the bands. We see that pairs of gapless magnon edge states appear in the vicinity of bulk gap signifying the strong topology of the system, yielding nonzero  $\kappa_{xy}$ \cite{me}. At the saturation field $h=h_s$ the spins are collinear along the $z$-axis corresponding to  bilayer ferromagnet coupled ferromagnetically. The DMI again leads to gap magnon bulk bands with counter-propagating gapless edge states and again with nonzero  $\kappa_{xy}$.

For the in-plane DMI $\boldsymbol{\mathcal D}_{ij}=\mathcal D{\hat{\bold x}}$ shown in Fig.~\ref{mo2}, the situation is different. There is no magnon Hall effect and   $\kappa_{xy}$ vanishes in all regimes \cite{me}, but there is topological magnon insulator as we now explain.  Indeed,   we have degenerate  magnon bulk bands between $S_z\to S_x=\pm S$ sectors at zero field, but the DMI  has a profound effect since the spins are along the $x$-$y$ Kagom\'e plane.    As shown in Fig.~\ref{mo2}(a) there is a pair of edge state modes for each spin sector, and they are related by time-reversal symmetry. This is an evidence of topological magnon insulator. However,  $\kappa_{xy}$ vanishes as a consequence of time-reversal symmetry between the degenerate spin sectors. In fact, this system is a magnonic counterpart to  fermionic  topological insulator with imaginary second-nearest-neighbour SOC between electron spin up and down \cite{guo}. For $h<h_s$ the bands cross at ${\pm\bf K}$ as shown above due to staggered flux configurations and the edge modes are not topologically protected as we have confirmed by computing the Berry curvatures and the Chern numbers. For this reason $\kappa_{xy}$ again vanishes.  At the saturation field $h=h_s$ the in-plane DMI disappears in the noninteracting limit and the system is topologically trivial with vanishing  $\kappa_{xy}$. The key observation in the layer antiferromagnetic system is that although topologically protected edge states are present at zero magnetic magnetic field,  $\kappa_{xy}$ is suppressed by antiferromagnetic coupling. 

Now, let us consider Model II which corresponds to the Kagom\'e material KFe$_3$(OH)$_{6}$(SO$_{4}$)$_2$ \cite{sup1,sup1a,sup2}. As shown above this model differs significantly from Model I due to the presence of spin scalar chirality which survives in the absence of DMI and magnetic ordering.  The associated  magnon edge modes are depicted in Fig.~\ref{mo3}. At zero field $h=0$ it is evident that there are no protected chiral edge modes. This shows that the system is topologically trivial and  $\kappa_{xy}$ vanishes for any values of DMI. In the presence of a magnetic field perpendicular to the Kagom\'e plane there is an induced noncoplanar spin texture which provides spin scalar chirality  \cite{sup1a}. Figures~ \ref{mo3}(b) and (c) show that the system is topologically nontrivial in this regime with protected gapless edge states which yield nonzero Chern number and finite  $\kappa_{xy}$ even without the presence of DMI \cite{me}. Indeed, the presence of spin scalar chirality is the basis of chiral spin liquid physics, therefore it will not be surprising that the nontrivial topology of this system persists in the spin liquid phase of the frustrated Kagom\'e magnets. Model III explicitly ignores the DMI and the easy-plane anisotropy  stabilizes the $\bold q=0$ magnetic ordering. This system is analogous to Model II as shown in Fig.~\ref{mo4}. This is because the presence of the DMI does not have any topological effects in frustrated Kagom\'e lattice unlike insulating Kagom\'e ferromagnets.  The layer antiferromagnetic system and the frustrated system have similarities and differences.  In both systems   $\kappa_{xy}$  vanishes at zero field which can be attributed to zero net magnetic moment. In other words,  the degeneracy at zero field between $S_z\to S_x=\pm S$ sectors in layer antiferromagnetic system yields a zero net magnetic moment and for the coplanar/non-collinear system at zero field we have $\sum_{\Delta}{\bf S}_{\Delta}=0$ on each triangular plaquette which also yields a zero net magnetic moment. However, the origin of finite $\kappa_{xy}$ is different in both systems. For the layer antiferromagnets, topological magnon bands is induced by the DMI, whereas in the frustrated system with coplanar/non-collinear ordering the concept of topological magnon bands originates from field-induced spin scalar chirality which is nonzero even in the absence of DMI and magnetic ordering $\la {\bf S}_j\ra=0$.

\section{Discussion and Conclusion}
It is natural to ask the importance  of this investigation and whether such nontrivial topological effects can be   experimentally realized in insulating antiferromagnets. Recently, topological magnon insulator has been realized in the Kagom\'e ferromagnet Cu(1-3, bdc) \cite{alex5a}. This material is also the first Kagom\'e ferromagnet that shows magnon Hall effect  with finite  $\kappa_{xy}$ \cite{alex6}. A recent experiment has reported a finite  $\kappa_{xy}$ in frustrated Kagom\'e volborthite Cu$_3$V$_2$O$_7$(OH)$_2$ $\cdot$2H$_2$O in the presence of an out-of-plane magnetic field $h=15~\text{Tesla}$ \cite{wat}.  This result is attributed to spin excitations in the spin liquid  regime. As mentioned previously, the Kagom\'e  volborthite is known to exhibit different magnetic-field-induced ordered phases for $h< 15 ~\text{Tesla}$ \cite{Yo,Yo1}, and the frustrated Kagom\'e compound Ca$_{10}$Cr$_7$O$_{28}$ \cite{Balz} also exhibits ferromagnetic ordered states for $h\sim 11 ~\text{Tesla}$. This suggests that the observed low temperature dependence of $\kappa_{xy}$ in Kagom\'e  volborthite might not be due to spin excitations in the spin liquid regime, but magnon excitations in the field-induced ordered phases.  The iron jarosite KFe$_3$(OH)$_{6}$(SO$_{4}$)$_2$ \cite{sup1a,sup2}  is an ideal  Kagom\'e antiferromagnet with $\bold q=0$ ground state and nonzero field-induced spin scalar chirality \cite{sup1a}. This is a perfect material to search for topologically nontrivial excitations with finite  $\kappa_{xy}$ as described in the present study and Ref.~\cite{me}. At the moment,  inelastic neutron scattering experiment has not figured out how to measure magnon edge state modes, which are consequences of the magnon bulk topology that gives rise to finite  $\kappa_{xy}$ because it is a bulk sensitive method. The magnon edge modes can be probed by edge sensitive methods such as  light \cite{luuk} or electronic \cite{kha} scattering method.  This is not an impossible task in principle, and we believe it will be measured in the near future.
\section*{Acknowledgement}
 Research at Perimeter Institute is supported by the Government of Canada through Industry Canada and by the Province of Ontario through the Ministry of Research
and Innovation.

\end{document}